\documentclass[bjps]{imsart}
\usepackage{amsthm,amsmath}
\usepackage[round]{natbib}
\numberwithin{equation}{section}
\startlocaldefs
\newcommand{\qued}{\rule{2mm}{3.5mm}}
\endlocaldefs
\begin{document}
\begin{frontmatter}
\title{A Note on the Berman Condition}
\runtitle{Berman Condition}

\begin{aug}
\author{\fnms{Rolf} \snm{Turner}\thanksref{a,t1}\ead[label=e1]%
{r.turner@auckland.ac.nz}}
\and
\author{\fnms{Patrick} \snm{Chareka}\thanksref{b}}
\thankstext{t1}{
Adjunct Professor in the Department of
Mathematics and Statistics at the University of New Brunswick,
Fredericton, N.B., Canada}
\affiliation[a]{Starpath Project, University of Auckland}
\affiliation[b]{St. Francis Xavier University}
\address[a]{Starpath Project\\
University of Auckland\\
Private Bag 92019\\
Auckland, New Zealand 1142\\
\printead{e1}}
\address[b]{
Department of Mathematics, Statistics\\
and Computer Science\\
St. Francis Xavier University\\
Antigonish, N. S., Canada B2G 2W5}
\runauthor{Turner \& Chareka}
\end{aug}

\begin{abstract}
It is established that if a time series satisfies the Berman
condition, and another related (summability) condition, the result
of filtering that series through a certain type of filter also
satisfies the two conditions.  In particular it follows that if
$X_t$ satisfies the two conditions and if $X_t$ and $a_t$
are related by an invertible ARMA model, then the $a_t$
satisfy the two conditions.
\end{abstract}



\end{frontmatter}

\section{Introduction}
\label{sec:intro}
The condition (on the autocovariances $\gamma_k$ of a stationary time series)
\begin{equation}
\lim_{k \rightarrow \infty} |\gamma_k| \; \ln k = 0
\label{eq:berman}
\end{equation}
was introduced by \citet[Theorem 3.1, page 510]{ber}.  It appears to
have been adopted as a fundamental sufficient condition in proving
results about extreme value distributions for correlated data.  It is
cited for instance in \citet[2.5.1, p.~444]{lead1}, \citet[5.1,
p.~248]{lin}, \citet[4.1.1, p.~80]{lead2}, \citet[Theorem 3.8.2,
p.~169; see also p.~198]{gal}, and in \citet[Theorem 4.4.8,
p.~217]{emb}, where it is described as being ``very weak''.  It
appears to be effectively the weakest condition that one can assume
and still obtain positive results in this context.

In \citet*{cha} the authors found it necessary to assume, in addition
to the Berman condition, another condition
\begin{equation}
\sum_{k=1}^{\infty} \frac{|\gamma_k|}{k^{\epsilon}} < \infty
\mbox{~~for some~~} \epsilon < 1.
\label{eq:sum.cond}
\end{equation}
This is given as condition (7) on page 598 of \citeauthor{cha}.  In that
paper the authors find it expedient to deal with the residuals from
fitting an ARMA model to the time series $X_t$ under consideration.
They are thereby concerned with the innovation terms of such a model.
Suppose that $X_t$ and $a_t$ are related via an ARMA model in which
the $a_t$ play the role of the innovations.  \citeauthor{cha} remark
that if the time series $X_t$ is a fractionally integrated ARMA
(``FARIMA'') time series (whence it satisfies the two conditions
of interest (\ref{eq:berman}) and (\ref{eq:sum.cond})) then the
innovations $a_t$ also form a FARIMA series provided that the
model is invertible.  Hence the $a_t$ satisfy the two conditions
of interest as well.

\citeauthor{cha} assert that more is true:  if $X_t$ is \emph{any}
stationary time series satisfying (\ref{eq:berman}) and
(\ref{eq:sum.cond}) and if $X_t$ and a series of innovations $a_t$
are related by an invertible ARMA model, then the $a_t$ will also
satisfy these conditions.  In this note we present the proof of
that claim.

We now remark that interest is focussed on the $a_t$ and these
quantities are thought of as being the output of a filter, with
the $X_t$ being the input.  However the phrasing of the claim,
with the $a_t$ being the innovations of an ARMA model, makes it
appear as if the $a_t$ are the \emph{input} to a filter, which is
rather confusing.  The required condition of invertibility
of the ARMA model is also somewhat disconcerting.  Finally,
it turns out that a slightly stronger claim may be established.
We therefore re-phrase the assertion to be proven, in a stronger
and less confusing form, and state the original claim as a corollary
of the re-phrased assertion.

\section{The Main Result}
\label{sec:result}

We state the result to be proven as follows:\\

\noindent
{\bf Theorem:} Suppose that $X_t$ is a stationary time series with
autocovariances $\gamma_k$ satisfying conditions (\ref{eq:berman})
and (\ref{eq:sum.cond}) and that the series $Y_t$ is the output of
a linear filter with input $X_t$ given as follows:
\[
Y_t = \sum_{n=0}^{\infty} \psi_n X_{t-n}
\]
Suppose that the $\psi_n$ are summable (whence the $Y_t$ form a
stationary time series).  Furthermore suppose that the $\psi_n$
satisfy the condition
\begin{equation}
|\gamma^W_k| \leq C r^{|k|} \mbox{~~for all~~} k
\label{eq:bound}
\end{equation}
for some constants $C$ and $r$, $0 < r < 1$, where
\[
\gamma^W_k = \sum_{n = -\infty}^{\infty} \psi_n \psi_{n+k}
\]
and where we set $\psi_n = 0$ for $n < 0$ (to simplify the notation).
Then the autocovariances $\gamma^Y_k$ of the series $Y_t$ satisfy
(\ref{eq:berman}) and (\ref{eq:sum.cond}) as well.\\

\noindent {\bf Proof:}\\
\begin{quote}

We remark that the $\gamma^W_k$ are in fact the autocovariances of
a time series $W_t$ defined by
\[
W_t = \sum_{n=0}^{\infty} \psi_n b_{t-n}
\]
where $b_t$ is white noise with variance 1.

Observe that
\begin{eqnarray*}
\gamma^Y_k & = & \sum_{n=-\infty}^{\infty} \sum_{m=-\infty}^{\infty}
                 \psi_n \psi_m \gamma_{m-n+k} \\
           & = & \sum_{h=-\infty}^{\infty} \sum_{n=-\infty}^{\infty}
                 \psi_n \psi_{n+h} \gamma_{k+h} \\
           & = & \sum_{h=-\infty}^{\infty} \gamma^W_h \gamma_{k+h}
\end{eqnarray*}

To show that the $\gamma^Y_k$ satisfy condition
(\ref{eq:berman}) we write
\begin{eqnarray*}
\gamma^Y_k
 & = & \sum_{h=-\infty}^{-k-1} \gamma^W_h \gamma_{k+h} +
                 \sum_{h=-k}^{-1} \gamma^W_h \gamma_{k+h} +
                 \sum_{h=0}^\infty \gamma^W_h \gamma_{k+h} \\
 & = & \sum_{j=1}^{\infty} \gamma^W_{k+j} \gamma_j +
       \sum_{j=0}^{k-1} \gamma^W_{k-j} \gamma_j +
       \sum_{j=0}^{\infty} \gamma^W_j \gamma_{k+j} \\
 & = & \xi_1(k) + \xi_2(k) + \xi_3(k) \mbox{~~(say).}
\end{eqnarray*}

To deal with $\xi_1(k)$ we observe that
\[
|\xi_1(k)| \ln k
\leq \sum_{j=1}^{\infty} |\gamma^W_{k+j}| |\gamma_j| \; \ln k
\leq C \; \gamma_0 \; r^k \ln k \; \frac{r}{1-r}
\]
using (\ref{eq:bound}).  This quantity $\rightarrow 0$
as $k \rightarrow \infty$ since $r < 1$.

Similarly
\[
|\xi_3(k)| \ln k \leq 
\sum_{j=0}^{\infty} |\gamma^W_j| |\gamma_{k+j}| \ln (k+j) \; \leq
C \sum_{j=0}^{\infty} r^j |\gamma_{k+j}| \ln (k+j)
\]
Take $\delta > 0$; for sufficiently large $k$, $|\gamma_{k+j}| \ln (k+j)
\leq \delta$ for all $j \geq 0$.  Hence
\[
|\xi_3(k)| \ln k \leq \frac{\delta \times C}{1-r}
\]
for sufficiently large $k$, and since $\delta$ is arbitrary,
$ |\xi_3(k)| \ln k \rightarrow 0 $ as $k \rightarrow \infty$.

To deal with the middle term $\xi_2(k)$ we note that
\begin{eqnarray*}
|\xi_2(k)| & \leq & C \sum_{j=0}^{k-1} |\gamma_j| r^{k-j} \\
 & = & C \left [ \sum_{j=0}^{[k/2]} |\gamma_j| r^{k-j}
               + \sum_{j=[k/2]+1}^{k-1} |\gamma_j| r^{k-j}
         \right ] \\
 & \leq & C \sum_{j=0}^{[k/2]} \gamma_0 r^{k-j}
               + C \sum_{j=[k/2]+1}^{k-1} |\gamma_j| r^{k-j} \\
 & \leq & C \gamma_0 \frac{r^{k/2}}{1-r}
               + C \gamma_{j^*(k)} \frac{r}{1-r}
\end{eqnarray*}
where $j^*(k) = \mbox{argmax} \{|\gamma_j| : [k/2]+1 \leq j \leq k-1 \}$.

Hence
\begin{eqnarray*}
\ln k \times |\xi_2(k)| & \leq & C \left [ \gamma_0 \ln k \frac{r^{k/2}}{1-r}
                            + (\ln k/2 + \ln 2) \left ( \gamma_{j^*(k)}
                               \frac{r}{1-r} \right ) \right ] \\
 & \leq & C \left [ \gamma_0 \ln k \frac{r^{k/2}}{1-r}
                            + (\ln j^*(k) + \ln 2) \left ( \gamma_{j^*(k)}
                               \frac{r}{1-r} \right ) \right ] \\
\end{eqnarray*}
which $\rightarrow 0$ as $k \rightarrow \infty$.

We have thus established that the autocovariances $\gamma^Y_k$
satisfy (\ref{eq:berman}).  We now proceed to show that condition
(\ref{eq:sum.cond}) is satisfied:

\begin{eqnarray*}
\sum_{k=1}^{\infty} \frac{|\gamma^Y_k|}{k^{\epsilon}}
 & = & \sum_{k=1}^{\infty} \left | \sum_{h=-\infty}^{\infty} \gamma^W_h
       \frac{\gamma_{k+h}}{k^{\epsilon}} \right |
       \leq \sum_{h=-\infty}^{\infty} |\gamma^W_h| \sum_{k=1}^{\infty}
       \frac{|\gamma_{k+h}|}{k^{\epsilon}} \\
 & = & \sum_{h=-\infty}^{-1} |\gamma^W_h| \sum_{k=1}^{\infty}
       \frac{|\gamma_{k+h}|}{k^{\epsilon}} +
       \sum_{h=0}^{\infty} |\gamma^W_h| \sum_{k=1}^{\infty}
       \frac{|\gamma_{k+h}|}{k^{\epsilon}} \\
 & = & \sum_{j=1}^{\infty} |\gamma^W_j| \sum_{k=1}^{\infty}
       \frac{|\gamma_{k-j}|}{k^{\epsilon}} +
       \sum_{h=0}^{\infty} |\gamma^W_h| \sum_{k=1}^{\infty}
       \frac{|\gamma_{k+h}|}{k^{\epsilon}} \\
 & \leq & \sum_{j=1}^{\infty} |\gamma^W_j| \sum_{k=1}^{\infty}
       \frac{|\gamma_{k-j}|}{k^{\epsilon}} +
       \sum_{h=0}^{\infty} |\gamma^W_h| \sum_{k=1}^{\infty}
       \frac{|\gamma_{k+h}|}{(k+h)^{\epsilon}}
       \left (\frac{k+h}{k}\right)^{\epsilon} \\
 & \leq & \sum_{j=1}^{\infty} |\gamma^W_j| \sum_{k=1}^{\infty}
       \frac{|\gamma_{k-j}|}{k^{\epsilon}} +
       \zeta \sum_{h=0}^{\infty} |\gamma^W_h| (1+h)
       \mbox{, where~~} \zeta = \sum_{k=1}^{\infty}
       \frac{|\gamma_{k}|}{k^{\epsilon}}
\end{eqnarray*}
\begin{eqnarray*}
\phantom{\sum_{k=1}^{\infty} \frac{|\gamma^Y_k|}{k^{\epsilon}}}
 & = & \sum_{j=1}^{\infty} |\gamma^W_j| \left [ \sum_{k=1}^{j}
       \frac{|\gamma_{k-j}|}{k^{\epsilon}} + \sum_{k=j+1}^{\infty}
       \frac{|\gamma_{k-j}|}{k^{\epsilon}} \right ] +
       \zeta \sum_{h=0}^{\infty} |\gamma^W_h| (1+h) \\
 & \leq & \sum_{j=1}^{\infty} C r^j \left [ \gamma_0 \sum_{k=1}^{j}
       \frac{1}{k^{\epsilon}} +
       \sum_{\ell=1}^{\infty} \frac{|\gamma_{\ell}|}{\ell^{\epsilon}}
       \left ( \frac{\ell}{\ell+j} \right)^\epsilon
       \right ] + \zeta \sum_{h=0}^{\infty} C r^h \; (1+h) \\
 & \leq & C \sum_{j=1}^{\infty} r^j \left [ j \gamma_0 +
       \sum_{\ell=1}^{\infty} \frac{|\gamma_{\ell}|}{\ell^{\epsilon}}
       \right ] + \zeta C \sum_{h=0}^{\infty} r^h \;(1+h) \\
 & \leq & C \sum_{j=1}^{\infty} r^j [ j \gamma_0 + \zeta] +
            \zeta C \sum_{h=0}^{\infty} r^h \;(1+h) \\
 & = & C \left [ \gamma_0 \sum_{j=1}^{\infty} jr^j +
         \zeta \sum_{j=1}^{\infty}  r^j + \zeta \sum_{j=0}^{\infty} r^j
         + \zeta \sum_{j=1}^{\infty} j r^j \right ] \\
 & = & C \left [ (\gamma_0 + \zeta) \sum_{j=1}^{\infty} j r^j +
       \zeta \; \frac{1+r}{1-r} \right ] < \infty
\end{eqnarray*}
since $\sum_{j=1}^{\infty} jr^j$ converges.  (The radius of convergence
of this power series is 1, and by assumption $0 < r < 1$.)
\hfill \qued\\
\end{quote}

\section{The Original Claim}
\label{sec:orig.claim}

We state the original claim as:\\

\noindent
{\bf Corollary:}  Suppose that $X_t$ satisfies conditions
(\ref{eq:berman}) and (\ref{eq:sum.cond}) and that $X_t$ and $a_t$
are related by the invertible ARMA model
\[
\phi(B)X_t = \theta(B) a_t
\]
where $\phi(z)$ and $\theta(z)$ are polynomials and $B$ is the
``backshift'' operator.  Then the $a_t$ satisfy conditions
(\ref{eq:berman}) and (\ref{eq:sum.cond}) as well.\\

\noindent {\bf Proof:}\\
\begin{quote}
The invertibility of the model tells us that $a_t$ can be
expressed as
\[
a_t = \sum_{n=0}^{\infty} \psi_n X_{t-n}
\]
where
\[
\frac{\phi(z)}{\theta(z)} = \psi(x) = \sum_{n=0}^{\infty} \psi_n z^n
\]
with the coefficients $\psi_n$ being summable.  (It is more usual in
such a context to denote the coefficients of the series expansion
as ``$\pi_n$'' rather than ``$\psi_n$'', but to make clear the
relationship to the main result we eschew the $\pi_n$ notation.)

Now if we set
\[
W_t = \sum_{n=0}^{\infty} \psi_n b_{t-n}
\]
where $b_t$ is white noise (with variance 1) then basic results
about ARMA time series (see e.g. \citet[Chapter 3, problem 3.11]{bro}) tell
us that the autocovariances $\gamma^W_k$ of $W_t$ satisfy
(\ref{eq:bound}).  Hence the claim follows by the theorem proven
in section~\ref{sec:result}.\\
\mbox{ } \hfill \qued\\
\end{quote}

\section*{Acknowledgments}
The authors express their warm thanks to an anonymous referee whose
comments substantially improved the exposition in this paper.  The
first author's research is supported by a grant from the Natural
Sciences and Engineering Research Council of Canada.

\bibliographystyle{melvin}
\bibliography{nbc}
\end{document}